\documentclass[12pt]{iopart}
\usepackage{graphicx}
\usepackage[hidelinks, colorlinks=True, citecolor=blue ]{hyperref}
\usepackage[dvipsnames]{xcolor}
\usepackage{cite}

\begin{document}

\title[Preface]{Preface: Characterisation of Physical Processes from Anomalous Diffusion Data}

\author{Carlo Manzo$^\dagger$, Gorka Mu\~noz-Gil$^\ddagger$, Giovanni
Volpe$^\sharp$, Miguel Angel Garcia-March$^\flat$, Maciej Lewenstein$^{\S}$, and Ralf Metzler$^\P$}

\address{$\dagger$ Facultat de Ci\`encies, Tecnologia i Enginyeries,
Universitat de Vic -- Universitat Central de Catalunya (UVic-UCC), C. de la
Laura,13, 08500 Vic, Spain\\
$\ddagger$ Institute for Theoretical Physics, University of Innsbruck, Technikerstr. 21a, A-6020 Innsbruck, Austria \\
$\sharp$ Department of Physics, University of Gothenburg, Origov{\"a}gen
6B, SE-41296 Gothenburg, Sweden\\
$\flat$ Instituto Universitario de Matem\'{a}tica Pura y Aplicada,
Universitat Polit\`{e}cnica de Val\`{e}ncia, Spain\\
$\S$
ICFO -- Institut de Ci\`encies Fot\`oniques, The Barcelona
Institute of Science and Technology, Av. Carl Friedrich Gauss 3, 08860
Castelldefels (Barcelona), Spain\\ and
ICREA, Pg. Llu\'is Companys 23, 08010 Barcelona, Spain\\
$\P$
Institute of Physics \& Astronomy, University of Potsdam,
Karl-Liebknecht-St. 24/25, D-14476 Potsdam-Golm, Germany\\
and Asia Pacific Centre for Theoretical Physics, Pohang 37673, Republic
of Korea}
\eads{\mailto{carlo.manzo@uvic.cat}, \mailto{rmetzler@uni-potsdam.de}}


%
\maketitle

\section*{Introduction}
Since Albert Einstein provided a theoretical foundation~\cite{einstein1905molekularkinetischen} for Robert Brown's
observation of the movement of microscopic granules contained in pollen
grains~\cite{brown1828brief}, significant deviations from the laws of Brownian
motion have been uncovered in an impressively wide variety of animate and
inanimate systems, from biology to the stock market. Anomalous diffusion,
as it has come to be called, extends the concept of Brownian motion and is
connected to disordered systems, non-equilibrium phenomena, flows of energy
and information, and transport in living systems~\cite{metzler2014anomalous}.
Anomalous diffusion is ``non-universal'' in the sense that physically
very different systems share the same power-law form of the mean squared
displacement $\langle x^2(t)\rangle\sim t^\alpha$. To properly understand
a system exhibiting anomalous diffusion, it is therefore important to have
reliable analysis methods to unveil the exact physical mechanisms effecting
the anomalous diffusion dynamics.

Several methods for detecting the occurrence of and the mechanisms
behind anomalous diffusion have been developed using classical
statistics~\cite{thapa2018bayesian, burnecki2015estimating,
weron2017ergodicity,sikora2017elucidating, kepten2015guidelines,
krapf2019spectral, thapa2020leveraging}. However, in the last years,
the booming of machine learning has boosted the development of data-driven
methods to characterise anomalous diffusion from single trajectories, providing
more refined tools for this problem~\cite{munoz2020single, granik2019single,
bo2019measurement, kowalek2019classification}.

In 2020, we launched the Anomalous Diffusion (AnDi) Challenge to
provide the first assessment of classical and novel methods for
quantifying anomalous diffusion in various realistic conditions through
a community-based effort~\cite{munoz2020anomalous}. The challenge
consisted of an open competition to benchmark existing methods and spur
the invention of new approaches. The AnDi Challenge brought together
a vibrating and multidisciplinary community of scientists working on
this problem, involving more than 30 participants from 22 institutions
and 11 countries. Ultimately, the analysis of the results obtained
on a reference dataset~\cite{munoz2020dataset} provided an objective
assessment of the performance of methods to characterise anomalous
diffusion from single trajectories for three specific tasks, including
anomalous diffusion exponent inference, model classification, and trajectory
segmentation. The study, published in Nature
Communications, analyses the results of the community effort and determines
that machine learning greatly improves the estimation of the properties of
diffusing particles~\cite{munoz-gil2021objective}.

This special issue includes the details of several of the methods that
participated in the AnDi challenge. Some of the articles describe updated versions of the software originally used for the challenge, showing improved performance. Most of these methods rely on state-of-the-art machine learning approaches. For instance, Ref.~\cite{gentili2021characterization} combines feature engineering based on classical statistics with feed-forward neural networks. Interestingly, this work shows how to create an adapted
pipeline specific to the dataset of the challenge to
reach some of the best performance of the competition across all tasks. Similarly,
Ref.~\cite{kowalek2022boosting} also uses a set of statistical
features as the input to an extreme gradient boosting model, which then takes care of classifying trajectories among diffusion models.  The authors further show that the proper choice of features
heavily affects classification accuracy.

Among the different machine learning approaches, recurrent neural networks (RNN) have attracted a lot
of interest due to their suitability when dealing with data with temporal
information and long-range correlations. Three works show different implementations based on RNN for the challenge tasks~\cite{garibo2021efficient, argun2021classification, li2021wavenet}. Ref.~\cite{garibo2021efficient} combines a bidirectional long short-term memory (LSTM), a state-of-the-art RNN, with a convolutional neural network (CNN). The CNN is used as a
feature extractor before a stack of LSTM layers to boost the RNN performance, achieving outstanding results for the regression of the anomalous exponent. Ref.~\cite{argun2021classification} proposes an architecture where trajectories are directly fed to LSTM layers, an implementation that enables the analysis of time traces of arbitrary size, without the need for any padding or preprocessing. This method shows top performance across several tasks of the AnDi challenge, demonstrating that similar architectures can be successfully used for different purposes. Last, Ref.~\cite{li2021wavenet} describes
the use of one of the most promising RNN architectures, the WaveNet. This work further shows how the size of the training dataset (one order of magnitude larger than the rest of the models used in the challenge) can be key to enhancing the method's accuracy.

Other ML approaches have also shown remarkable performance in dealing with
stochastic diffusion. Ref.~\cite{manzo2021extreme} attempts to establish a baseline for machine learning approaches using extreme learning machine (ELM), a fast-converging training algorithm for single hidden layer feedforward neural network applied over a set of statistical features. Ref.~\cite{al2022classification} uses a pretrained CNN, the ResNet-50, to classify different stochastic processes and compare it with other CNNs. 
Ref.~\cite{conejero2022characterization} proposes a new architecture, the convolutional Transformer, to extract features from trajectories and feed them to two transformer encoding blocks that perform either regression or classification. Ref.~\cite{verdier2021learning} presents a method based on graph neural networks (GNN) where a vector of features is associated with each trajectory position and a sparse graph structure with each trajectory. Similar to Ref.~\cite{li2021wavenet}, the authors use representation learning techniques to study the latent space features of their model and propose a visual exploratory method to analyse trajectories from walks never seen by the GNN. In fact, the unsupervised analysis of diffusion models is a promising tool to characterize unknown datasets that could even lead to the identification of new mechanisms. Along this line, Ref.~\cite{munoz2021unsupervised} studies the suitability of auto-encoders as feature extractors for anomalous diffusion trajectories and proposes a method to characterise them using anomaly detection.

Besides machine learning approaches, theory-based methods were also proposed for characterising anomalous diffusion and tested in the AnDi challenge. Ref.~\cite{meyer2022decomposing} discusses numerical methods to obtain the anomalous diffusion exponent and proposed a questionnaire for model selection based on feature analysis. Bayesian inference was instead used in Ref.~\cite{thapa2022bayesian} to distinguish between scaled and fractional Brownian motion and in Ref.~\cite{park2021bayesian} that presents an approach to deal with L\'evy walk trajectories. Bayesian methods are particularly effective when enough information is known about the trajectories and specific priors associated with the type of walk one aims to characterise can be constructed. 

The special issue also hosts several theoretical contributions pushing
forward the field of stochastic processes and/or discussing applications to
time series. Thus, Ref.~\cite{vitali2022anomalous} discusses emerging
transient anomalous diffusion in Markovian hopping-trap scenarios. Transient
anomalous diffusion is also obtained in a tempered fractionally integrated
process~\cite{sabzikar2022tempered}. Ref.~\cite{maraj2020empirical}
introduces the empirical anomaly measure as a means to measure the distance
between the anomalous diffusion process and normal diffusion. Limit properties
of L{\'e}vy walks are shown to be useful in the recognition and verification of
L{\'e}vy walk-type motion, as well as the parameter estimation in maximum
likelihood methods~\cite{magdziarz2020limit}. Ref.~\cite{wang2020fractional}
studies the emerging residual nonergodicity in fractional Brownian motion with
random diffusivity that may help distinguish and categorise certain
nonergodic and non-Gaussian features of particle displacements. Applications
of single-trajectory power spectral methods to movement data of kites and
storks are discussed in Ref.~\cite{vilk2022classification}, demonstrating
how stochastic models can be extracted with this method.

Quantum walks are considered in Ref.~\cite{hegde2022characterization}, showing
how the interplay between quantum coherence and the mean squared displacement
of the walker can provide information on the process.
Ref.~\cite{ablowitz2022integrable} studies applications of the inverse
scattering transform to fractional versions of non-linear equations of,
e.g., the Korteweg-deVries equation, that provide a framework for solitonic
solutions with power-law dispersion relations. Initial strong non-Gaussianity
concurrent with Brownian scaling of the mean squared displacement is reported
for self-avoiding random walks in Ref.~\cite{baldovin2022}.

The research reported in this special issue provides a major contribution
toward the understanding of anomalous diffusion processes and their
analysis. In particular, a palette of tools is introduced, which are poised
to become standard methods for the analysis of trajectories generated from
various experiments, from atomic physics to ecology. Moreover, the outcome
of these studies reinforces the importance of community-based efforts in
the search for the advancement of science. The success of this initiative
triggered us to organise the ${\rm 2^{nd}}$ AnDi Challenge around the problem
of detecting changes in transport properties and interactions between moving
objects from single trajectories.
\section*{Acknowledgments}
We thank all the authors for their contributions, the editorial staff of
the journal for their support, and the participants of the AnDi Challenge for their involvement.\\
\\
\sloppy
CM acknowledges support through the grant RYC-2015-17896 funded by MCIN/AEI/10.13039/501100011033 and ``ESF Investing in your future'', grants BFU2017-85693-R and PID2021-125386NB-I00 funded by MCIN/AEI/10.13039/501100011033/ and ``ERDF A way of making Europe''.
GMG acknowledges support from the European Union (ERC, QuantAI, Project No. 10105529) and the Austrian Science Fund (FWF) through the SFB BeyondC F7102.
GV acknowledges funding from ERC StG ComplexSwimmers (Grant No. 677511) and from the Knut and Alice Wallenberg Foundation.
MAGM acknowledges funding from the Spanish Ministry of Education and Professional Training (MEFP) through the Beatriz Galindo program 2018 (BEAGAL18/00203), Spanish Ministry MINECO (FIDEUA PID2019-106901GBI00/10.13039/501100011033), QuantERA II Cofund 2021 PCI2022-133004, Project of MCIN with funding from European Union NextGenerationEU (PRTR-C17.I1) and by Generalitat Valenciana, with Ref. 20220883 (PerovsQuTe).
ML acknowledge support from: ERC AdG NOQIA; Ministerio de Ciencia y Innovaci\'on Agencia Estatal de Investigaciones (PGC2018-097027-B-I00/10.13039/501100011033,  CEX2019-000910-S/10.13039/501100011033, Plan National FIDEUA PID2019-106901GB-I00, FPI, QUANTERA MAQS PCI2019-111828-2, QUANTERA DYNAMITE PCI2022-132919,  Proyectos de I+D+I “Retos Colaboración” QUSPIN RTC2019-007196-7); MICIIN with funding from European Union NextGenerationEU(PRTR-C17.I1) and by Generalitat de Catalunya;  Fundaci\'o Cellex; Fundaci\'o Mir-Puig; Generalitat de Catalunya (European Social Fund FEDER and CERCA program, AGAUR Grant No. 2017 SGR 134, QuantumCAT/U16-011424, co-funded by ERDF Operational Program of Catalonia 2014-2020); Barcelona Supercomputing Center MareNostrum (FI-2022-1-0042); EU Horizon 2020 FET-OPEN OPTOlogic (Grant No. 899794); EU Horizon Europe Program (Grant Agreement 101080086 — NeQST), National Science Centre, Poland (Symfonia Grant No. 2016/20/W/ST4/00314); ICFO Internal “QuantumGaudi” project; European Union’s Horizon 2020 research and innovation program under the Marie-Skłodowska-Curie grant agreement No. 101029393 (STREDCH) and No. 847648  (``La Caixa'' Junior Leaders fellowships ID100010434: LCF/BQ/PI19/11690013, LCF/BQ/PI20/11760031,  LCF/BQ/PR20/11770012, LCF/BQ/PR21/11840013). Views and opinions expressed in this work are, however, those of the author(s) only and do not necessarily reflect those of the European Union, European Climate, Infrastructure and Environment Executive Agency (CINEA), nor any other granting authority.  Neither the European Union nor any granting authority can be held responsible for them.
RM acknowledges the German Science Foundation (DFG grant ME 1535/12-1)
for funding.

\section*{References}
\providecommand{\noopsort}[1]{}\providecommand{\singleletter}[1]{#1}%
\providecommand{\newblock}{}


\end{document}